# Quantum Hall Drag of Exciton Superfluid in Graphene

Xiaomeng Liu[1], Kenji Watanabe[2], Takashi Taniguchi[2], Bertrand I. Halperin[1], Philip Kim[1]

[1]*Department of Physics, Harvard University, Cambridge, Massachusetts 02138, USA*

[2]*National Institute for Material Science, 1-1 Namiki, Tsukuba 305-0044, Japan*

**Excitons are pairs of electrons and holes bound together by the Coulomb interaction. At low temperatures, excitons can form a Bose-Einstein condensate (BEC), enabling macroscopic phase coherence and superfluidity[1,2]. An electronic double layer (EDL), in which two parallel conducting layers are separated by an insulator, is an ideal platform to realize a stable exciton BEC. In an EDL under strong magnetic fields, electron-like and hole-like quasi-particles from partially filled Landau levels (LLs) bind into excitons and condense[3–11]. However, in semiconducting double quantum wells, this magnetic-field-induced exciton BEC has been observed only in sub-Kelvin temperatures due to the relatively strong dielectric screening and large separation of the EDL[8]. Here we report exciton condensation in bilayer graphene EDL separated by a few atomic layers of hexagonal boron nitride (hBN). Driving current in one graphene layer generates a quantized Hall voltage in the other layer, signifying coherent superfluid exciton transport[4,8]. Owing to the strong Coulomb coupling across the atomically thin dielectric, we find that quantum Hall drag in graphene appears at a temperature an order of magnitude higher than previously observed in GaAs EDL. The wide-range tunability of densities and displacement fields enables exploration of a rich phase diagram of BEC across Landau levels with different filling factors and internal quantum degrees of freedom. The observed robust exciton superfluidity opens up opportunities to investigate various quantum phases of the exciton BEC and design novel electronic devices based on dissipationless transport.**

An exciton BEC is formed when a large fraction of excitons occupy the ground state, establishing macroscopic coherence with weak dipolar repulsion[1,3]. In order to stabilize the exciton BEC, however, we need to prevent the recombination of electrons and holes, which leads to the annihilation of excitons. One way to achieve a large density of long-lived excitons is to place electrons and holes in spatially-separated parallel conducting layers, where excitons can form across the layers. In semiconducting EDLs, such indirect excitons can be formed by optical excitation[12] or electrical doping[13]. One salient feature of the exciton BEC is dissipationless exciton transport, consisting of counter-flowing electrical currents carried by co-traveling electrons and holes[4]. The first experimental observation of this superfluid exciton flow was demonstrated in GaAs EDLs under a strong magnetic field, in which a strong correlation is formed between electron-like and hole-like quasi-particles in quantizing orbits[3–11].

The magnetic-field-induced layer coherence of the EDL can be established in the following way. When a two-dimensional (2D) electron gas of density $n$ is subject to a perpendicular magnetic field $B$, the kinetic energy of electrons is quantized to discrete LLs. Each LL contains $n_0 = \frac{eB}{h}$ degenerate Landau orbits per unit area, where $e$ is electron charge and $h$ is Planck's constant. If all the orbits in a LL are occupied (i.e., the filling factor $v = n/n_0$ is an integer), the 2D electron system forms a quantum Hall state. In the EDL, the filling factor of the individual layer can be specified by $v_{top} = n_{top}/n_0$ and $v_{bot} = n_{bot}/n_0$, where $n_{top}$ and $n_{bot}$ are the density of top and bottom layer, respectively. If LLs in both layers are partially filled, i.e., $v_{top}$ and $v_{bot}$ are non-integer numbers, neither layer can form a quantum Hall state on its own. However, Coulomb repulsion forces the electrons in the two layers to occupy different orbitals in space, leading to spatial anti-correlation between layers. Notably, when the total filling fraction, $v_{tot} = v_{top} + v_{bot}$, becomes an integer, the two layers together can form a coherent state in which each filled state (quasi-electron) in the one layer correspond to an empty state (quasi-hole) in the other layer. These bound empty-filled states can be described as excitons in the quantum Hall scenario and yield a strong response in the Coulomb drag experiment: driving current in one layer generates a quantized Hall drag voltage in the other layer. To be specific, the ratio between drive current ($I^{drive}$) and drag Hall voltage ($V_{xy}^{drag}$), defined as drag Hall conductivity, is then quantized as $\sigma_{xy}^{Drag} = I^{drive}/V_{xy}^{drag} = \frac{e^2}{h} v_{tot}$. Meanwhile, Hall conductivity of the drive layer ($\sigma_{xy}^{Drive}$), obtained from the drive Hall voltage ($V_{xy}^{drive}$), is also quantized to the same value despite having a partially filled LL, while the longitudinal voltage in both layers vanishes.

Quantized Hall drag for $v_{tot} = 1$ has been observed in the lowest LLs in GaAs EDLs[8]. Coherent tunneling between the layers[7] and perfect current drag measurements[11] further confirmed the presence of the interlayer coherence and the superfluid exciton flow. The BEC realized in semiconducting EDLs, however, turns out to be rather weak, with a BEC transition temperature $T_c$, limted to the sub-Kelvin range. This fragility is mainly caused by a relatively large EDL separation $d$. Loosely-confined electron wave functions in semiconducting EDL typically require $d > 20$ nm to prevent directly tunneling between the conducting layers and to avoid the interfacial disorder effect. It is noted that $T_c$ is proportional to the characteristic energy scale $e^2/\varepsilon l$, where $\varepsilon$ is the dielectric constant and $\ell = (\hbar/eB)^{1/2}$ is the magnetic length specifying the distance between quasi-particles in a LL[14,15]. Also, the exciton BEC only appears in the strong coupling regime[4], where the $d/\ell$ ratio is below the critical value of $d/\ell < 2$. Thus, together with small $\varepsilon$ of the graphene EDL, reducing $d$ substantially below the limit of the semiconducting EDL will likely enhance $T_c$ and also increase the exciton binding energy.

Recent progress on 2D van der Waals heterostructures has created a new opportunity to build EDLs using atomically thin materials[16–19]. Owing to electron-hole symmetry and extremely light carrier mass, EDLs consisting of mono- and bilayer graphene have been of particular interest to realize an exciton BEC[20–23]. Furthermore, tunneling currents in graphene-hBN-graphene heterostructures are not appreciable when $d > 1.5$ nm at small biases, due to the large bandgap of hBN and tightly bound electron wave function in graphene[16,24]. Initial experiments performed in graphene-hBN-graphene heterostructures demonstrated a strong Coulomb drag effect in the semiclassical regime realized at high temperatures, exhibiting the strong interaction between the two layers in the zero and finite magnetic fields[16–18]. However, experimental evidence of interlayer coherence has yet to be found. In this work, with improved device quality and fabrication technique (see supplementary information (SI)), we demonstrate magnetic-field-induced exciton condensation in graphene EDL.

Our devices are made of two Bernal stacked bilayer graphene sheets separated by 3 nm hBN and encapsulated by two thicker hBN layers (20~30 nm) (Fig. 1a,b). The two graphene layers are independently contacted by multiple electrodes (see SI for details). Both graphene layers have mobility $0.5~1 \times 10^6 \text{cm}^2/\text{Vs}$ and exhibit symmetry breaking quantum Hall states at fields as low as 5 T (SI, Fig. S3-4). No appreciable tunneling current is measured above the noise level, providing a lower bound on the tunneling resistance of 1 GΩ. The voltages applied to the top gate ($V_{TG}$), the bottom gate ($V_{TG}$) and the interlayer bias between graphene layers ($V_{int}$) tune the carrier densities of the top and bottom graphene layers ($n_{top}, n_{bot}$): $n_{top} = C_{TG}V_{TG} -$ $C_{int}V_{int}$ ; $n_{bot} = C_{BG}V_{BG} + C_{int}V_{int}$. Here $C_{TG}$, $C_{BG}$, and $C_{int}$, are capacitances between the top gate and top layer, the bottom gate and bottom layer, and between the top and bottom graphene layers, respectively. By controlling $V_{TG}$, $V_{BG}$, and $V_{int}$, we can also adjust the average displacement fields; $D_{top} = (C_{TG}V_{TG} + C_{int}V_{int})/2$ and $D_{bot} = (-C_{BG}V_{BG} + C_{int}V_{int})/2$, for the top and bottom layer, respectively. For the drag measurements, we apply current $I^{drive}$ only in the top layer and measure the Hall resistance $R_{xy}^{drive}$ of the current-carrying (drive) layer, and magneto- and Hall drag resistances, $R_{xx}^{drag}$ and $R_{xy}^{drag}$, of the drag layer under a perpendicular magnetic field $B$. Owing to the Onsager relation, switching the drive and drag layer produces experimentally equivalent drag results (SI, Fig. S5).

Fig. 1c shows measurements of $R_{xy}^{drive}, R_{xx}^{drag}, R_{xy}^{drag}$ under $B$ = 25 T, corresponding to the strong coupling limit ($\ell$ = 5.1 nm and $d/\ell$ = 0.58). In this plot, we adjust $V_{TG}$ and $V_{BG}$ such that the filling fractions of each layer are balanced ($\nu \equiv \nu_{drive} = \nu_{drag}$). We observe that each layer exhibits its own quantum Hall (QH) effect. For $\nu \geq 1$, $R_{xy}^{drive}$ exhibit QH plateaus at the values $(R_{xy}^{drive})^{-1} = \frac{e^2}{h}, \frac{4}{3}\frac{e^2}{h}, \frac{2e^2}{h}$. In these well-developed integer ($\nu$=1, 2) and fractional ($\nu$=4/3) QH regimes of the individual layers, we find no appreciable drag signal ($R_{xx}^{drag} \approx R_{xy}^{drag} \approx 0$). The vanishing drag signals at low temperatures are expected in the semiclassical picture due to the diminishing scattering phase space[25]. However, the observed drag signals are significantly enhanced when the first LL of both layers are partially filled ($\nu$ < 1). In particular, for $\nu$=1/2, where both layers are half-filled and thus $\nu_{tot} = \nu_{drive} + \nu_{drag} = 1$, the Hall drag signal reaches close to the quantization value of $h/e^2$ = 25.8 kΩ, while the magneto-drag ($R_{xx}^{drag}$) dips to nearly zero. Under the same condition, the Hall resistance in the drive (top) layer, which originally rises beyond $h/e^2$ as $\nu_{drive}$ drops below one (i.e., partially filled LL), re-enters $h/e^2$ again at $\nu_{tot}$= 1. This re-entrant behavior of $R_{xy}^{drive}$ to the same quantized value of $R_{xy}^{drag}$ indicates that the entire EDL behaves like a single $\nu$=1 quantum Hall system despite that LLs in each layer are only partially filled.

The quantized Hall drag and re-entrant QHE in the drive layer have been observed previously in the GaAs EDLs for $\nu_{tot}$= 1 and are considered as a strong evidence of interlayer coherence and exciton superfluidity[8]. The follow-up perfect drag measurement[11] further demonstrated the existence of dissipationless superfluid current flows in this condensate. A simple physical picture for the observed quantized Hall drag can be built upon a two-fluid picture. In this model, currents in each layer are carried by excitons in the bulk ($I_{ex}^{(i)}$) and quasi-particles flowing on the edge ($I_{qp}^{(i)}$), where the superscript is the layer index. Excitons generate counter flow currents $I_{ex}^{drag} = -I_{ex}^{drive}$; and the zero accelerating electric force requirement on superfluid excitons demands $V_{Hall}^{drive} = V_{Hall}^{drag} = V_{Hall}$. In addition, boundary conditions of the drag and drive layers requires $I_{ex}^{(drag)} + I_{qp}^{(drag)} = 0$, $I = I_{ex}^{drive} + I_{qp}^{drive}$. Furthermore, by considering the two layers as a single coherent quantum Hall system at filling fraction $v_{tot}$, we have $I_{qp} = I_{qp}^{(drag)} + I_{qp}^{(drive)} = \frac{v_{tot}e^2}{h}V_{Hall}$. Summing up, we obtain the experimental observation $R_{xy}^{drag} = R_{xy}^{drag} = h/v_{tot}e^2$ with vanishing $R_{xx}^{drag}$ and $R_{xx}^{drive}$.

We found that the observed quantized Hall drag in graphene is much more robust than that of the GaAs EDLs. In the insert of Fig. 1c, we show the temperature dependence of Hall drag at B = 13 T in the temperature range T=0.9 - 10K. The signature of the exciton condensate, i.e., simultaneous quantization of $R_{xy}^{drag}$ and $R_{xy}^{drag}$ to the value of $h/v_{tot}e^2$, persists up to a few Kelvin for $v_{tot} = 1$. Further quantitative analysis reveals the condensation energy gap $\Delta \approx$ 8 K using the activation formula $R_{xy}^{drive}(T) - h/e^2 \propto h/e^2 - R_{xy}^{drag}(T) \propto e^{-\Delta/T}$ (SI, Fig. S6b). Compared with the gap energy $\Delta \approx 0.8K$ in GaAs EDLs[8], a stabler exciton BEC is achieved in graphene EDLs with smaller layer separations.

The exciton BEC in graphene is also found to be robust against the density imbalance between layers. Fig. 1d-f displays $R_{xy}^{drag}$, $R_{xy}^{drag}$, and $R_{xy}^{drag}$ as a function of the $V_{TG}$ and $V_{BG}$ which tune $v_{drive}$ and $v_{drag}$, respectively. Interestingly, the signatures of the exciton condensation, i.e., $R_{xy}^{drag} \approx R_{xy}^{drag} \approx h/e^2$ and $R_{xx}^{drag} \approx R_{xx}^{drag} \approx 0$ withstand a range of gate voltages satisfying $v_{drag} + v_{drive} = 1$, corresponding to the diagonal line specified in each panel of the graphs. For a more quantitative analysis, we plot $R_{xy}^{drag}$ cut along this diagonal line as an overlay graph in Fig. 1d (white trace). The level of $R_{xy}^{drag}$ quantization indicates that the BEC persists for the density imbalance $\frac{\Delta n}{n_{tot}} = \frac{n_{drag}-n_{drive}}{n_{drag}+n_{drive}}$ up to $\sim \pm 30\%$. Beyond this limit the more stable integer QH states ($v_{drag}$ or $v_{drive} = 0$ or 1) in each layer take over the exciton BEC phase.

While the exciton BEC has been discovered only for the half-filled lowest LL in the GaAs EDLs, the gate tunability in graphene EDL devices allows us to explore the phase diagram of possible condensate states other than $v_{tot}= 1$ for a wide range of filling factors of the drive and drag layers. Fig. 2 shows experimental survey for $R_{xy}^{drag}$ as a function of $v_{drive}$ and $v_{drag}$, covering the electron-electron ($v_{drive}, v_{drag} > 0$) and hole-hole ($v_{drive}, v_{drag} < 0$) regimes. Remarkably, we find at least two additional exciton BEC states in these regimes: ($v_{drive}, v_{drag}$) centered near (0.5, 2.5) and (-1.5, -1.5), corresponding to the drag between ½ - 2½ filled electron LLs ($v_{tot} = 3$) and 1½ - 1½ filled hole LLs ($v_{tot} = -3$), respectively. Similar to the BEC in $v_{tot} = 1$, these states exhibit the quantized Hall drag; $R_{xy}^{drag} \approx h/v_{tot}e^2$ and $R_{xx}^{drag} \approx 0$, for a range of ($v_{drive}, v_{drag}$) satisfying $v_{drive} + v_{drag} = v_{tot}$ (see Fig 2. b-j). These quantized Hall drag features appear as diagonals in the ($v_{drive}, v_{drag}$) plots and are confined to the sectors corresponding to partially filled first (drive) and third (drag) electron LLs (with all symmetries are lifted) and partially filled second (drive and drag) hole LLs. Measurements at lower magnetic fields also reveal a signature of developing exciton BEC for ($v_{drive}, v_{drag}$)=(2.5, 0.5), the symmetric pair for (0.5, 2.5) discussed above (SI, Fig. S5&7). The relatively weak presence of this symmetric pair is presumably due to the quality difference between the top and bottom graphene layers. We also remark that while $R_{xy}^{drag} \approx h/v_{tot}e^2$ and $R_{xx}^{drive} \approx R_{xx}^{drag} \approx 0$ are observed in the $v_{tot} = \pm 3$ state, we find $|R_{xy}^{drive}| > h/v_{tot}e^2$ (Fig. 2e and k). We speculate that a dissipative exciton transport of a fragile BEC is responsible for this incomplete re-entrant QHE.

Interestingly, our experimental observations strongly indicate that the apparent electron-hole symmetry of LLs is broken for the exciton BEC. For example, the (0.5, 0.5) BEC exists while (-0.5, -0.5) is absent. Instead, the $v_{(i)} \rightarrow v_{(i)} + 2$ symmetry, which relate (-1.5, -1.5) to (0.5, 0.5) and (0.5, 0.5) to (0.5, 2.5) appears to hold. Similar $v \rightarrow v + 2$ symmetry has been observed in the filling fraction sequences of the fractional quantum Hall effect in bilayer graphene and related with the orbital degeneracy of bilayer graphene LLs[26].

We also note that the existence of the excitonic BEC at fixed ($v_{drive}, v_{drag}$) sensitively depends on the $V_{int}$ (Fig.3 and Fig. S7-9 in SI). For example, Fig. 3b shows $R_{xy}^{drag}$ as a function of $V_{int}$ while keeping $\nu_{tot} = 1$. Here, $D_{top} = (C_{int}V_{int} + n_{top})/2$ and $D_{bot} = (C_{int}V_{int} - n_{bot})/2$ are tuned by the interlayer bias $V_{int}$, while $V_{TG}$ and $V_{BG}$ are adjusted accordingly to keep $n_{top}$ and $n_{bot}$ unchanged. Since the displacement fields acting on individual bilayer graphene layers can tune the internal degree of freedoms of LL wave functions, tuning $V_{int}$ provides a tool to control the spin, valley and orbital texture of the interlayer exciton. Indeed, we observe that $R_{xy}^{drag}$ undergoes multiple distinct transitions between the quantized Hall value $h/e^2$ and 0 as $V_{int}$ changes, even at fixed $\nu_{tot} = 1$ as shown in Fig. 3b.

Close inspection indicates that the $V_{int}$ dependence of $R_{xy}^{drag}$ is intimately related to the transition of the symmetry-broken QH states of bilayer graphene in our EDL. Fig. 3a(c) depicts the $R_{xx}$ measured in the drive (drag) layer as a function of $V_{TG}$ ($V_{BG}$) and $V_{int}$. Phase transitions of different symmetry-broken QH states are clearly manifested. For example, while $R_{xx} \approx 0$ along most part of $\nu_{drive} = 1$ line (Fig. 3a), one can find three transition points where $R_{xx} > 0$, corresponding to $D^{top} = 0, \pm D_1$. These transition points were identified as phase transitions between partially layer-polarized and fully layer-polarized states in previous studies of generalized quantum Hall ferromagnetism (QHFM)[27,28]. We find the transitions in $R_{xy}^{Drag}$ coincide well with the phase transitions of $\nu = 1$ either in the drive or the drag layer. For example, the transition between region (1) and (2) in Fig. 3b align with $D^{top} = 0$ transition of the drive layer and the transition between region (2) and (3) matches with $D^{bot} = -D_1$ of the drag layer. More comprehensive analysis reveals that some transitions can also correspond to $\nu = 0$ transitions (orange dashed line in Fig. 3 and black dashed line in Fig. S9 in SI). The close connection between exciton BEC formation and the internal degrees of freedom of partially filled LLs suggests that these many body quantum states have a rich phase space to explore.

**Acknowledgements** We thank Amir Yacoby, Allan Macdonald, Andrea Young, and Laurel Anderson for helpful discussion. The major experimental work is supported by DOE (DE-SC0012260). The theoretical analysis was supported by the Science and Technology Center for Integrated Quantum Materials, NSF Grant No. DMR-1231319. P.K. acknowledges partial support from the Gordon and Betty Moore Foundation's EPiQS Initiative through Grant GBMF4543. K.W. and T.T. acknowledge support from the Elemental Strategy Initiative conducted by the MEXT, Japan. T.T. acknowledges support from a Grant-in-Aid for Scientic Research on Grant262480621 and on Innovative Areas \Nano Informatics" (Grant 25106006) from JSPS. A portion of this work was performed at the National High Magnetic Field Laboratory, which is supported by National Science Foundation Cooperative Agreement No. DMR-1157490 and the State of Florida. Nanofabrication was performed at the Center for Nanoscale Systems at Harvard, supported in part by an NSF NNIN award ECS-00335765.

**Author Contributions** XL performed the experiments and analyzed the data. XL and PK conceived the experiment. XL, BIH and PK wrote the paper. KW and TT provided hBN crystals.

Correspondence and requests for materials should be addressed P.K. (e-mail: pkim@physics.harvard.edu).

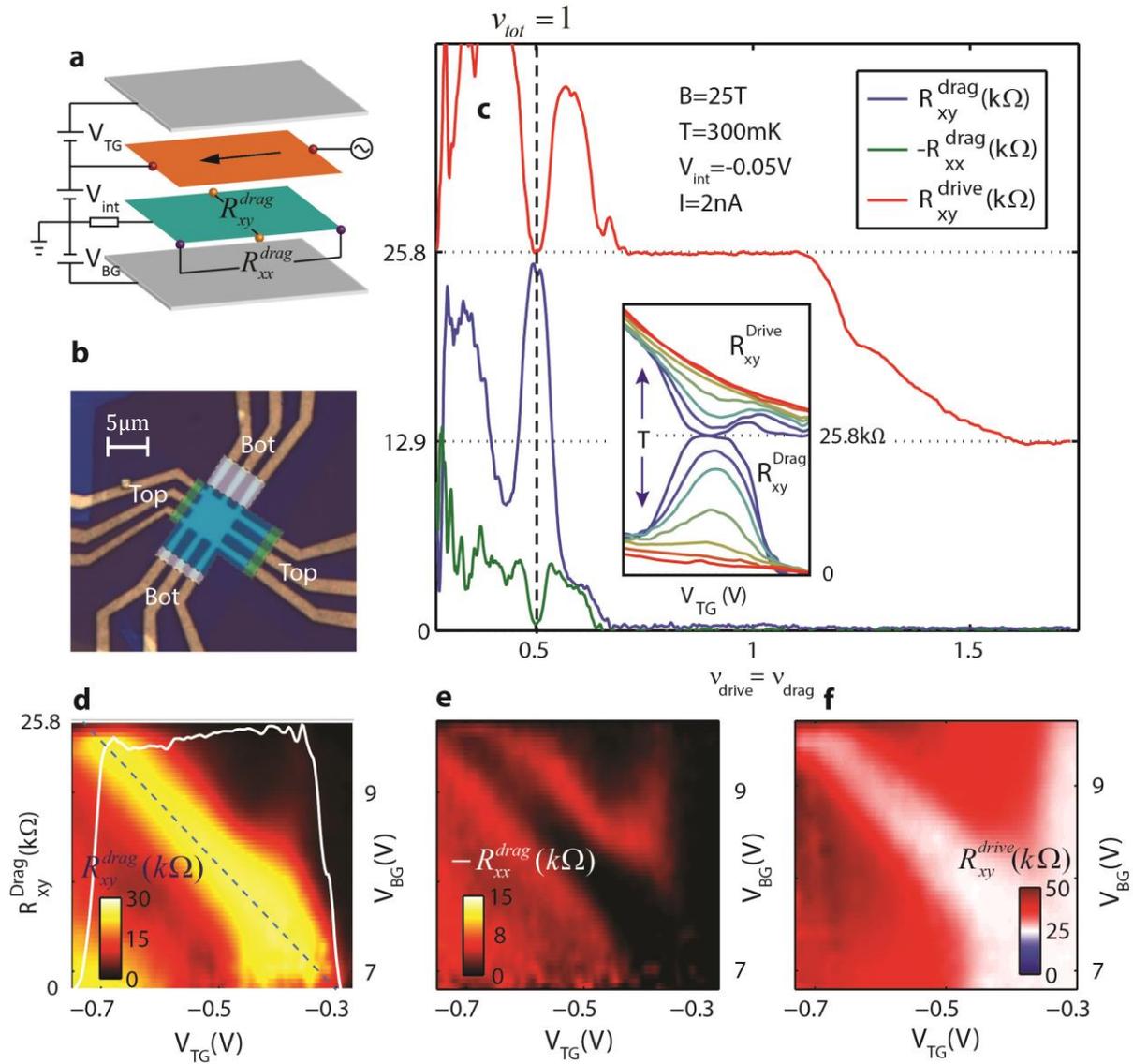

**Figure 1 | Quantized Hall drag for $v_{tot} = 1$ state in bilayer graphene double layers. a,** schematic diagram of device and measurement setup. **b**, Optical microscope image of the device. Metal leads on the left and right of the image (three on each side) contact the top layer graphene, while others contact the bottom layer graphene. The blue shaded area of graphene is under the top gate; white and green shaded regions are under the contact gate. The contact gates provides highly transparent electrical contacts under high magnetic fields. **c**, $R_{xy}^{drag}, R_{xx}^{drag}, R_{xy}^{drive}$ as a function of filling factors of both layers at *B*=25T and *T*=300mK. The exciton BEC can be recognized by quantized Hall drag ($R_{xy}^{drag} = \frac{h}{e^2}, R_{xx}^{drag} = 0$) with the simultaneous re-entrant quantum Hall in the drive layer. **c** insert, temperature dependence (*T*=0.9, 2.29, 3.13, 4.35, 6, 8, 10K) of $R_{xy}^{drag}$ at B=25T. **d, e, f**, $R_{xy}^{drag}, R_{xy}^{drive}, R_{xx}^{drag}$ as a function of $V_{TG}$ and $V_{BG}$ (y-axis labels on the right), tuning the filling fraction of each layer. The exciton BEC region appears as a diagonal region satisfying $v_{top} + v_{bot} = 1$. The white trace in **d** shows the value of $R_{xy}^{drag}$ (axis on the left) along $v_{tot} = 1$ line (dashed line in **d**).

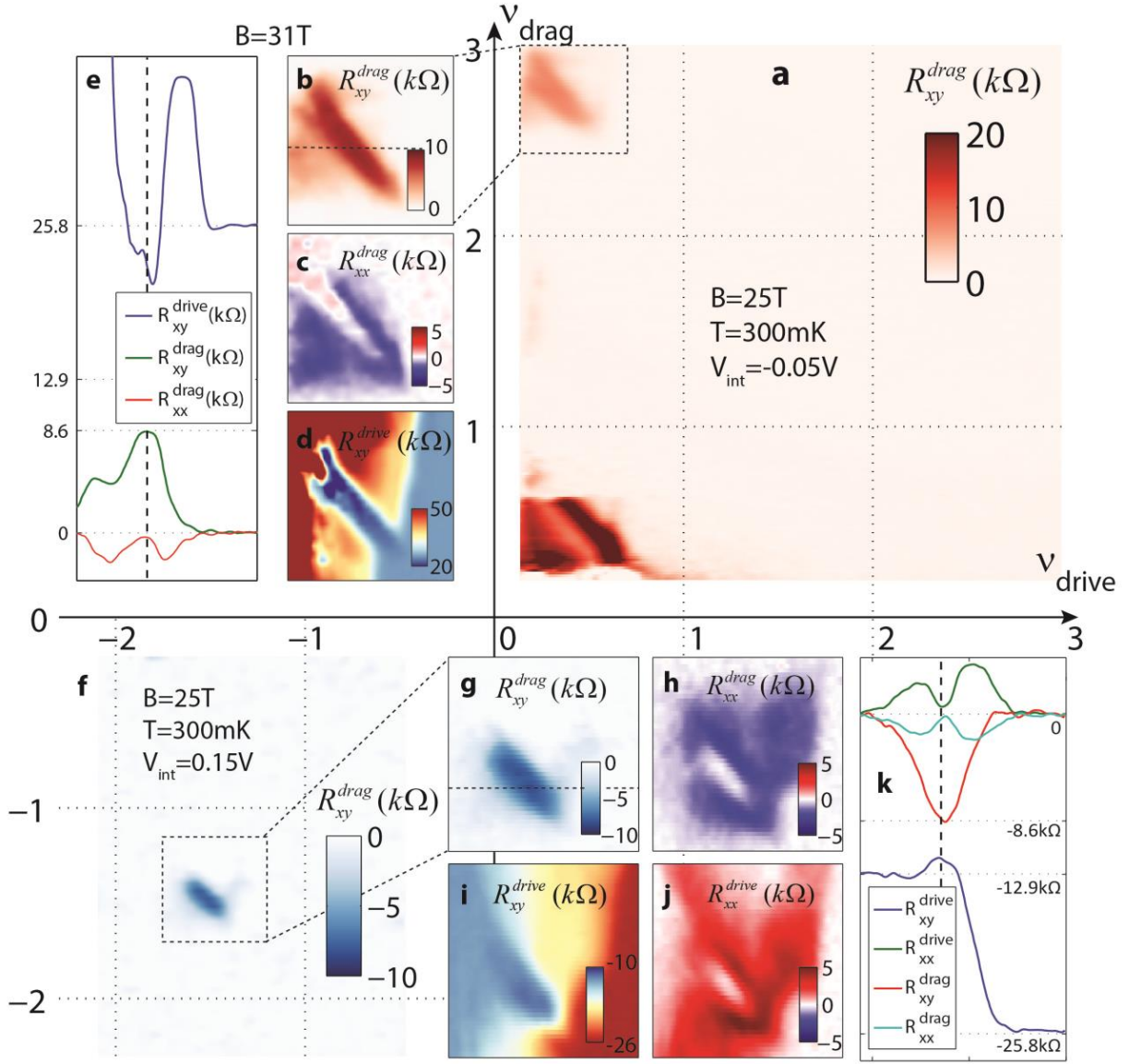

**Figure 2 | Exciton BEC in various LL fillings. a,** $R_{xy}^{drag}$ as a function of the top and bottom layer filling factors at B=25T, T=300mK and $V_{int}$= -0.05V. Besides ($\nu_{drag}$, $\nu_{drive}$) = (0.5, 0.5), additional exciton BECs are identified at the regions centered around (2.5, 0.5) and (-1.5, -1.5). **b, c, d,** zoomed-in plot of $R_{xy}^{drag}$, $R_{xx}^{drag}$, $R_{xy}^{drive}$ around $\nu_{tot} = 3$ at a higher field of B=31T. **e,** line-cut of $R_{xy}^{drag}$, $R_{xx}^{drag}$, $R_{xy}^{drive}$ along dashed line show in **b**. **f,** $R_{xy}^{drag}$ as function of filling factors at B=25T, T=300mK and $V_{int}$= 0.15V. **g, h, i, j,** $R_{xy}^{drag}$, $R_{xx}^{drag}$, $R_{xy}^{drive}$, $R_{xx}^{drive}$ at the same condition as **f**. **k,** line-cut of $R_{xy}^{drag}$, $R_{xx}^{drag}$, $R_{xy}^{drive}$, $R_{xx}^{drive}$ along the dashed line in **g**.

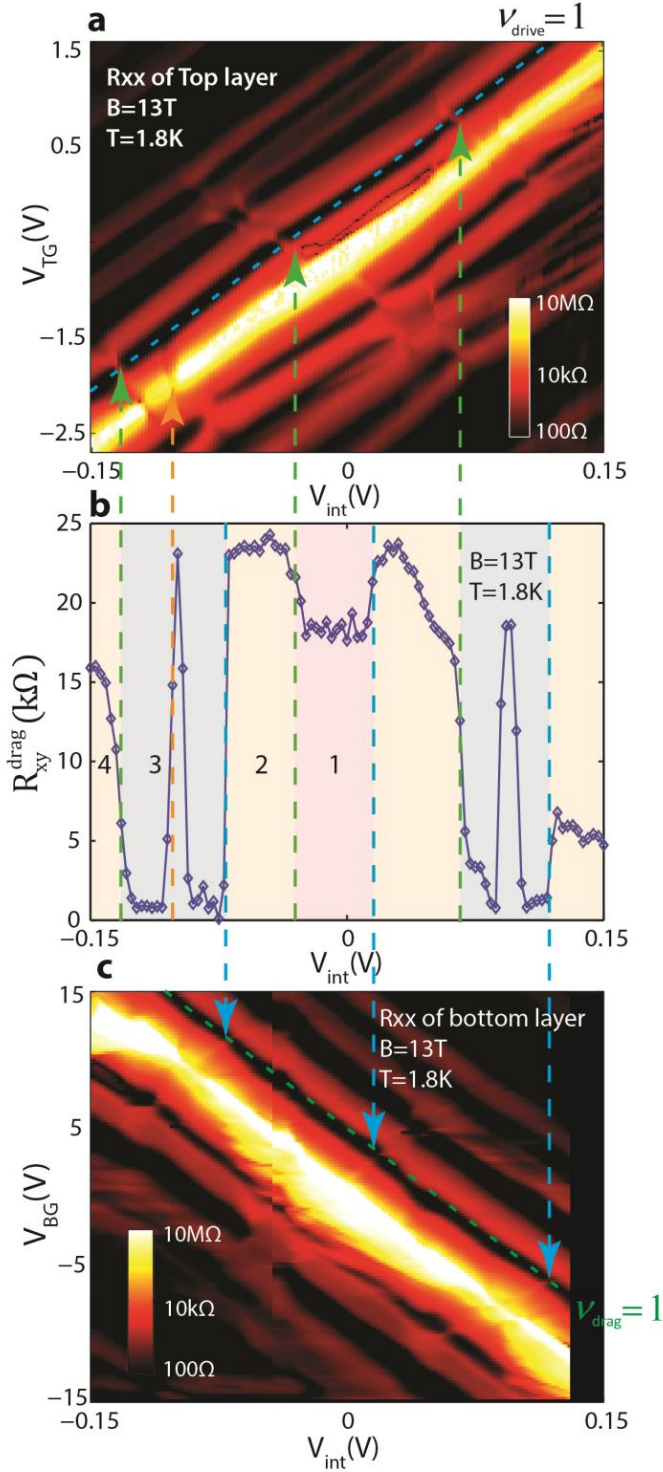

**Figure 3 | Phase transition of $v_{tot} = 1$ exciton BEC induced by transverse electric field. a**, $R_{xx}^{top}$ as a function of interlayer bias $V_{int}$ and top gate voltage $V_{TG}$ under $B$=13T and $T$=1.5K. The two voltages together effectively tune the carrier density and displacement field on the top layer. The dark regions ($R_{xx} \approx 0$) are where the quantum Hall effect sets in the top layer. The dashed line indicates where the density in the layer matches $v_{drive} = 1$. Along this line the displacement field on the top layer induces the quantum Hall ferromagnet transition between partial layer polarization to full layer polarization, marked by three green vertical arrows corresponding to $D^{top} = 0, \pm D_1$. **b**, $R_{xy}^{drag}$ as a function of interlayer bias $V_{int}$ while keeping $v_{tot} = 1$ at B=13T and T=1.5K. Tuning $V_{int}$ in this experiment effectively adjusts the perpendicular displacement fields in both drive and drag layers, staying at the same total filling fraction. Several example regions in $V_{int}$ where $R_{xy}^{drag}$ exhibits different behaviors are marked by (1) - (4). The fully quantized Hall drag signal is only stabilized in (2), whose phase boundaries in $V_{int}$ align with the transitions in $v = 1$ QHFM in the top layer (green up-arrows) and the bottom layer (blue down-arrows). A sharply peaked re-entrant quantized Hall drag only appears in (3) whose position aligns well with the transition points of top layer $v = 0$ QHFM (orange uparrow). **c**, $R_{xx}^{bottom}$ as a function of $V_{int}$ and bottom gate voltage $V_{BG}$ under same condition as in (a).

# Supplementary Information

## S1. Device fabrication

The five single-crystal layers of graphene and hBN (Fig. S1a) are prepared by mechanical exfoliation and van der Waals (vdW) transfer technique[1]. During the process, we choose two strips of bilayer graphene and align them into a cross, so we can use the overlapped part as the main channel area while fabricating individual contacts onto the non-overlapped parts (Fig. S1b). The edge contacts and the top gate are fabricated after etching the entire stack into the designed shape for the final device geometry. Usually, the parts of graphene near metal contacts are slightly doped and form unwanted PN junctions due to the metal-graphene work function mismatch. These contact-induced local PN junctions cause contact barriers under high magnetic fields. In order to address this problem, we fabricate local top gates (contact gates) to dope the area of graphene around the contact. These contact gates are adjusted to highly dope graphene leads next to the metal electrodes with the same carrier type of the channel (Fig. S1c).

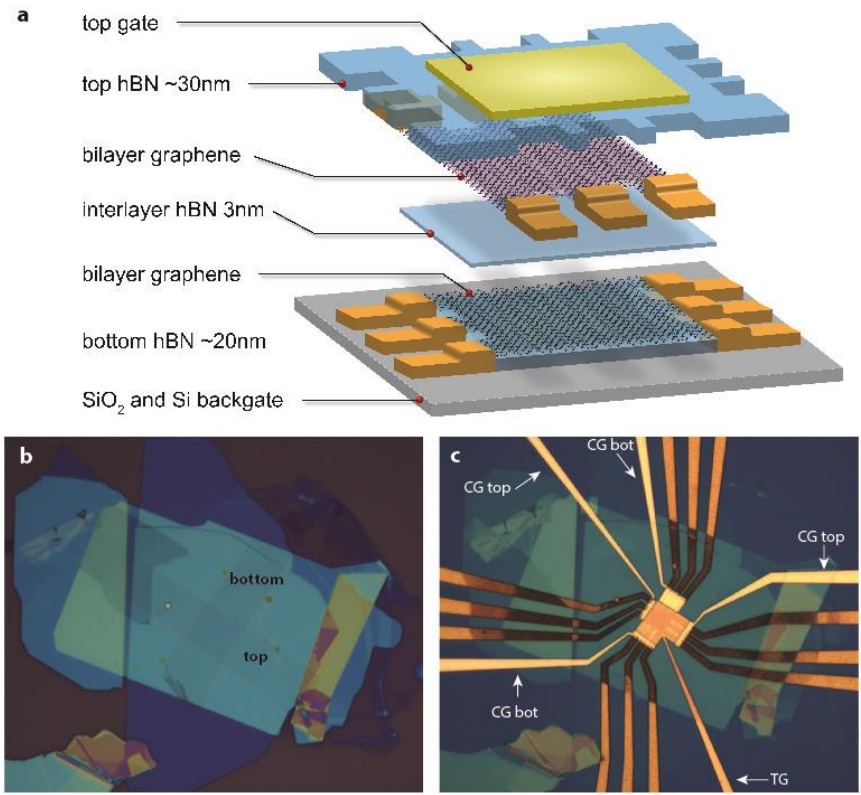

Fig. S1**a**, device structure. **b**, completed five-layer hetero-stack before contacts and gates fabrication. The dashed lines mark the boundary for the top and the bottom layer graphene. **c**, completed device image, the contact gates (CG) for the top and bottom graphene layers and the top gate (TG) are shown.

## S.2 Measurement setup

For the most part of our measurements, we use the top layer as the drive layer and the bottom layer as the drag layer (switching the drive and drag layers does not affect our drag experiment results as shown in Fig. S5). We apply $I^{drive}$ = 2nA AC current (17.7Hz) to the drive layer and measure the voltage drop along the longitudinal direction in the drag layer to obtain $V_{xx}^{drag}$. We also measure the voltage drops along the perpendicular direction in the drag and drive layers to obtain $V_{xy}^{drag}$ and $V_{xy}^{drive}$, respectively. The magneto and Hall drag resistances are obtained by $R_{xx}^{drag} = V_{xx}^{drag}/I^{drive}$ and $R_{xy}^{drag} = V_{xy}^{drag}/I^{drive}$ from this measurement. One of the contacts in the drag layer is connected to ground through a $1M\Omega$ resistor to allow the gates to tune the density of the drag layer.

A typical way to drive current in lock-in measurements is applying AC bias voltage $V^{drive}$ on one side of the graphene channel while grounding the other side. However, in Coulomb drag measurements, biasing the drive layer employing this scheme can create spurious drag signals in the drag layer due to drive bias induced AC gating. Because of the finite contact and channel resistances of the drive layer, the direct biasing scheme raises the potential in the middle of the drive layer to ~$V^{drive}/2$ with respect to ground. Since the drag layer is grounded, an AC interlayer bias of ~$V^{drive}/2$ is produced accordingly. In the previous drag experiments performed in GaAs double quantum wells, it has been shown that this AC interlayer bias induces spurious drag signals[2]. In order to prevent the AC interlayer bias, we employ a bridge setup[3] as shown in Fig. S2. In this scheme, the initial 4 Volts AC voltage is reduced down to 4mV through a voltage divider. Then this 4mV AC voltage is fed into the bridge through a 1:1 ground-isolating transformer. By tuning the variable resistor in the bridge, we can adjust the AC electrical potential in the middle of the drive layer to approximate zero. The 4mV is then converted into ~2nA by passing through two 1MOhm resistors in series. This setup also allows us to control the DC interlayer bias $V_{int}$. To monitor the drive current, we measure the voltage across one of the two 1MOhm resistors.

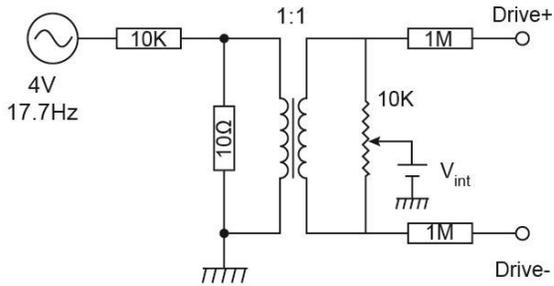

Fig. S2, Bias scheme for the drive layer.

### S.3 Quality of graphene channels

We measure the transport of properties of the drive and the drag layer independently at zero magnetic fields (shown in Fig. S3) where Hall mobility ~0.6 ×10$^6$cm$^2$/Vs was estimated at density ~10$^{11}$cm$^{-2}$ at 1.7 K.

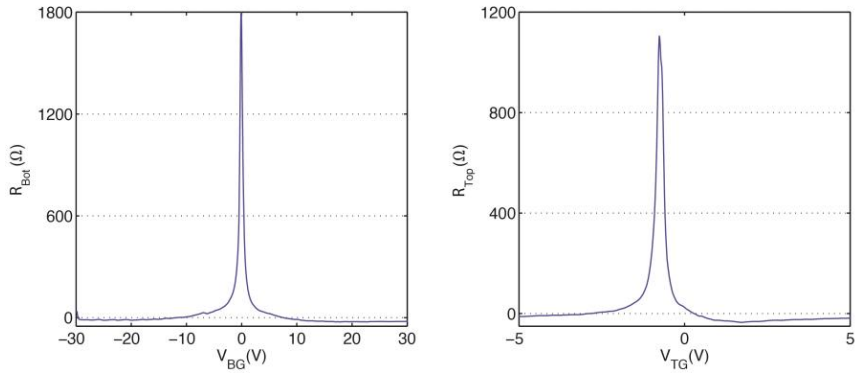

Fig. S3 Gate voltage dependent resistance measured in the top (a) and bottom layer (b) bilayer graphene used in this experiment.

High mobility and low disorders realized in the drive and drag layer channels enable observation of the symmetry broken quantum Hall states at a field as low as 5 T as shown in Fig. S4.

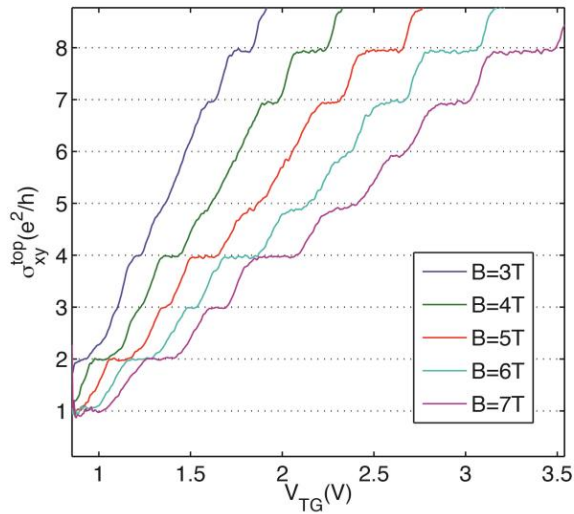

Fig. S4 Top layer Hall conductivity ($\sigma_{xy}^{top}$) as a function of top gate voltage under magnetic fields between 3 ~ 7 Tesla. The Hall conductivity plotted here is in the unit of conductance quantum $e^2/h$. At B=5T, most of the symmetry broken quantum Hall states are fully developed.

## S4. Onsager relation: drive and drag layer exchange

The Onsager reciprocity relation requires that changing the role of the drive and drag layers in our experiments produces identical results with reversal of the direction of magnetic field. Indeed, our experiment shows that exchanging the drive and drag layer measurement configuration yields nearly identical results under opposite magnetic field directions as shown in Fig. S5.

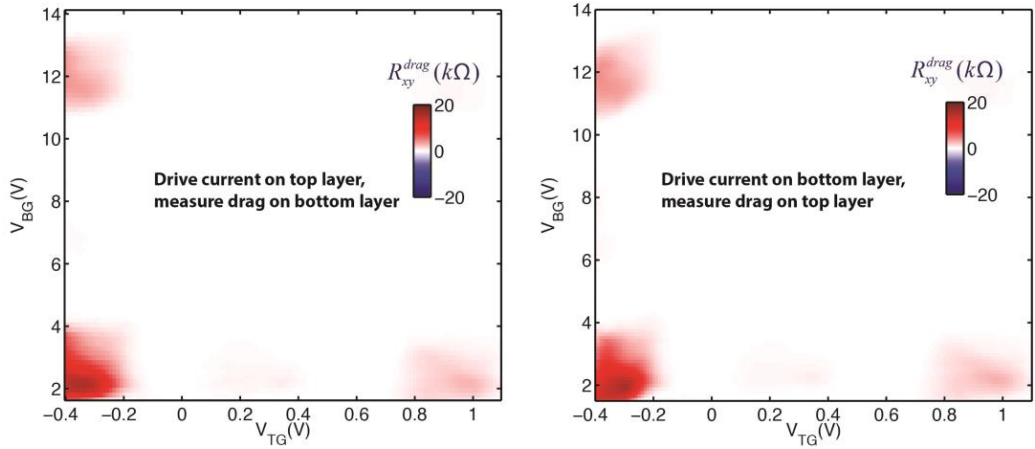

Fig. S5 Demonstration of Onsager reciprocity. Left panel shows drag measurements performed by driving the top layer and measuring the Hall drag voltage in the bottom layer at B=13T. Right panel shows a reciprocal configuration, i.e., driving the bottom layer and measuring the Hall drag voltage on the top layer (right panel) at B=-13T. Temperature is fixed to 1.5 K. Both experiments yield almost identical results except some small differences caused by different sets of contacts employed in the measurements.

## S5. Temperature dependence of $\nu_{tot} = 1$

As shown in Fig. S6a, the quantized Hall drag in our experiment becomes unstable as temperature increases. The exciton condensation energy $\Delta$ can be obtained from these temperature dependent behaviors of $R_{xy}^{drive}$ and $R_{xy}^{drag}$. Fig. S6b shows changes of the $R_{xy}^{drive}$ and $R_{xy}^{drag}$, defined as $\Delta R_{xy}^{drive} = R_{xy}^{drive} - h/e^2$, $\Delta R_{xy}^{drag} = h/e^2 - R_{xy}^{drag}$, as a function of temperature. In the same plot, we also display the minimum value of $R_{xx}^{drag}$, which is expected to vanish in a fully developed exciton BEC. As shown in this plot, all three quantities $\Delta R_{xy}^{drive}$, $\Delta R_{xy}^{drag}$, and $R_{xx}^{drag}$ exhibit an Arrhenius behavior $\sim e^{-\Delta/T}$ in the low temperature regime (T < 5 K), from which we estimate $\Delta \sim 8$ K.

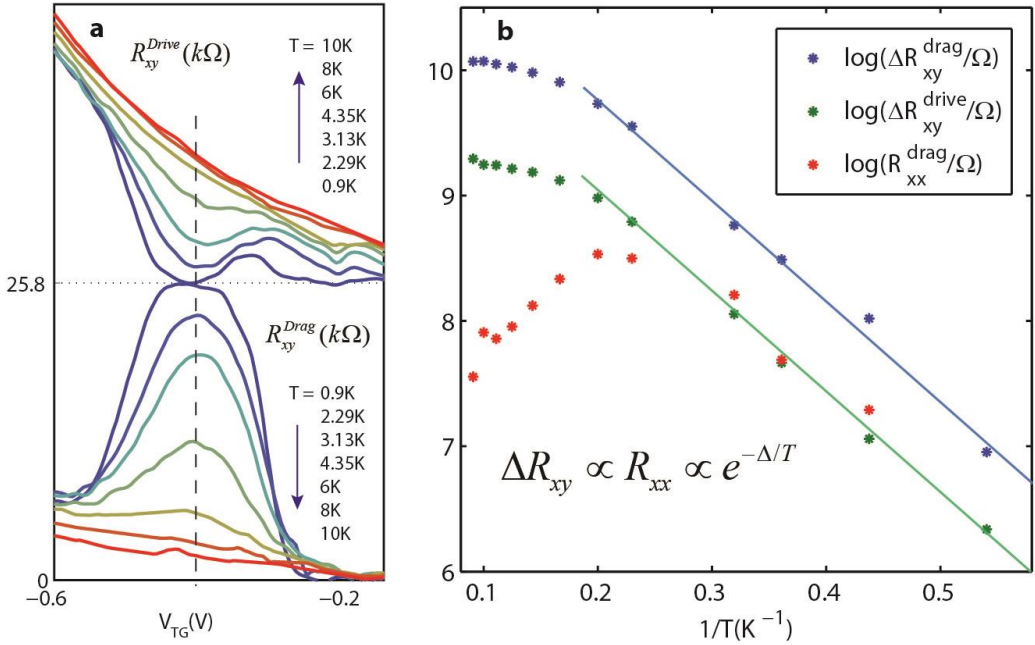

Fig. S6a, Hall resistance and Hall drag as a function of top gate at different temperatures at B=25T. b, $\Delta R_{xy}^{drive} = R_{xy}^{drive} - h/e^2$, $\Delta R_{xy}^{drag} = h/e^2 - R_{xy}^{drag}$ and $R_{xx}^{drag}$ as a function of 1/T at the gate voltage marked by vertical dashed line in a. The y-axis of this plot is in the log scale. Blue and green line in b marked an Arrhenius fit to $\Delta R_{xy}^{drag}$ and $\Delta R_{xy}^{drive}$ at the low temperature range from which we obtain the gap energy of ~8K.

## S6. Effect of $V_{int}$ on the Hall drag

The features of the exciton BEC in our experiment, such as quantized Hall drag and re-entrant QHE, sensitively depend on the applied interlayer bias $V_{int}$. Fig. S7 shows Hall drag as a function of $V_{TG}$ and $V_{BG}$ in the same filling factor range $0 < \nu^{drive}, \nu^{drag} < 3$, but at different interlayer biases. We notice at some interlayer biases, such as $V_{int}$ = -0.06V or 0.03V, strong Hall drag at $(\nu^{drive}, \nu^{drag}) = (0.5, 0.5), (0.5, 2.5), (2.5, 0.5)$ are manifested, while at some other $V_{int}$, such as $V_{int} = -0.12V$, no Hall drag is observed at all.

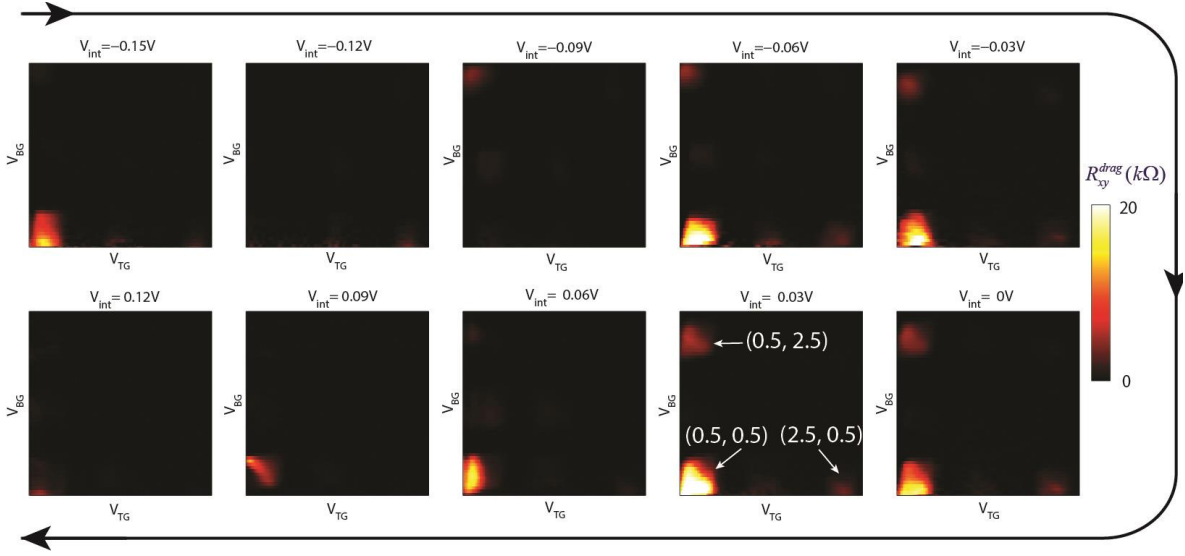

Fig. S7 Hall drag resistance as function of top and bottom gate at different interlayer biases at B=13T and T=1.5K in the electron-electron regime. The number marked in $V_{int}$ = 0.03V plot indicate $(\nu^{drive}, \nu^{drag})$.

A similar trend emerges in the hole-hole drag regime. Fig. S8 shows how Hall drag around $(\nu^{drive}, \nu^{drag}) = (-1.5, -1.5)$ evolves with $V_{int}$. The Hall drag is strongest at $V_{int} = 0.15V$ and completely vanishes at $V_{int} = 0.1V$ or $0.05V$.

Fig. S9 further demonstrates the relationship between the exciton BEC and the corresponding QHFM states of each layer, tuned by the interlayer bias. Here we plot $R_{xx}^{top}$ and $R_{xy}^{drag}$ as a function of $n^{top}$ and $D^{top}$. In Fig. S9a, three QHFM transition points of $\nu^{top} = 1$ on the top graphene layer can be identified by peaks in $R_{xx}^{top}$ (marked by the green arrows), while $R_{xx}^{top} = 0$ along other parts of $\nu^{top} = 1$ line. There are also four transition points marked by the black arrows corresponding to $\nu^{top} = 0$ QHFM transitions where $R_{xx}^{top}$ display a dip. Interestingly, we find drastic changes in the drag signal when the drive layer undergoes some of the QHFM phase transitions. In Fig. S9b, we plot corresponding Hall drag signals displaying several distinctive regions where clear transition points between them can be found. We attribute three out of six transition points to the top layer QHFM phase transitions. It is noted that at B=25T, two transition points in Hall drag correspond to $\nu^{top} = 0$ transitions (black dashed line) and one transition align with $\nu^{top} = 1$ transition (green dashed line) of the drive (top) layer.

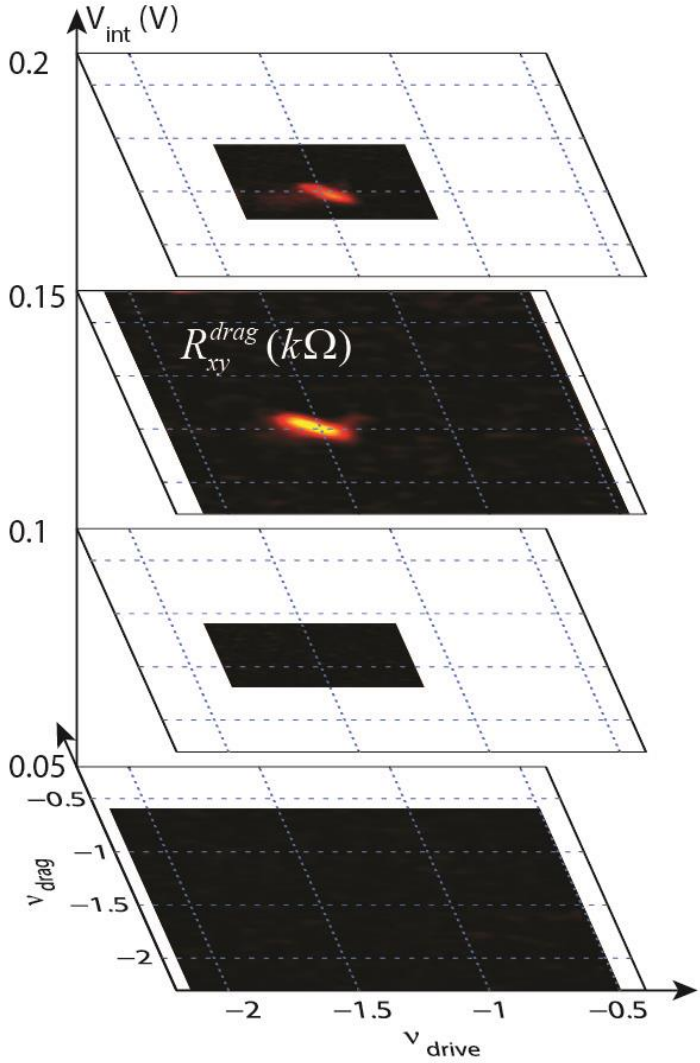

Fig. S8 Hall drag resistance as function of drive and drag filling factor at different interlayer bias at B=25T and T=0.3 K in the hole-hole drag regime for $V_{int}$=0.05V, 0.1V, 0.15V, 0.2V.

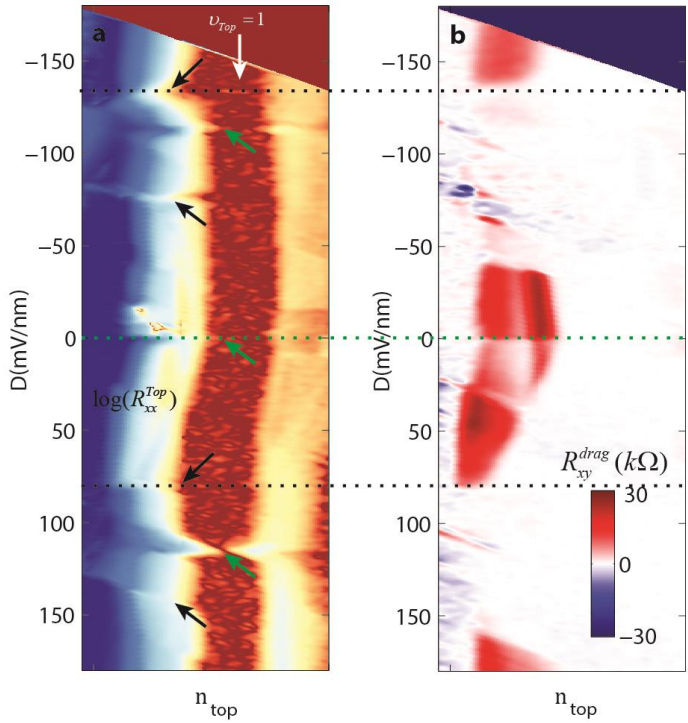

Fig. S9**a**, $R_{xx}^{top}$ as a function of density and displacement field of the top layer at B=25T and T=300mK. Green arrows indicate three QHFM phase transitions of $\nu = 1$ state and black arrow point to four QHFM phase transitions of $\nu = 0$ state. **b**, Hall drag ($R_{xy}^{drag}$) as a function of density and displacement field of the top layer. We noticed the behavior of $R_{xy}^{drag}$ exhibits multiple transitions in displacement field which can be aligned to $\nu = 1$ transition (green dashed line) or $\nu = 0$ transition (black dashed line) of the drive layer. The rest of the transition in $R_{xy}^{drag}$ is presumably correlated with the QHFM phase transtions in the drag layer.

### S.7 References